\def\'#1{\ifx#1i{\accent"13\i}\else{\accent"13#1}\fi}
\def\alamenos#1{$^{-#1}$}
\def\ala#1{$^{#1}$}

\def\kfor{{k_{\rm for}}}
\def\Ms{{M_{\rm s}}}

\def\tturb{\tau_{\rm turb}}

\documentclass[12pt,preprint]{aastex}

\usepackage{natbib}

\begin{document}

\title{The Mass Spectra of Cores in Turbulent Molecular Clouds and 
Implications for the Initial Mass Function}

\author{Javier Ballesteros-Paredes$^{1}$, Adriana Gazol\ala 1,
Jongsoo Kim\ala 2, Ralf S. Klessen\ala 3, Anne-Katharina Jappsen\ala
3, and Epimenio Tejero\ala 1}

\affil{$^1$Centro de Radioastronom\'ia y Astrof\'isica, UNAM,
Apdo. Postal 72-3 (Xangari), Morelia, Michoac\'an 58089, M\'exico}
\email{j.ballesteros, a.gazol@astrosmo.unam.mx}

\affil{$^2$ Korea Astronomy and Space Science Institute, 61-1,
Hwaam-Dong, Yuseong-Gu, Daejeon 305-348, Korea}
\email{jskim@kasi.re.kr}

\affil{$^3$Astrophysikalisches Institut Potsdam, An der Sternwarte 16,
  14482 Potsdam, Germany}
\email{rklessen@aip.de}

\slugcomment{Draft date: \today}

\lefthead{Ballesteros-Paredes et al. }

\righthead{Mass Spectra from Turbulent Fragmentation}

\begin{abstract}

  We investigate the core mass distribution (CMD) resulting from
  numerical models of turbulent fragmentation of molecular clouds. In
  particular we study its dependence on the sonic root-mean-square
  Mach number $\Ms$. We analyze simulations with $\Ms$ ranging from 1
  to 15 to show that, as $\Ms$ increases, the number of cores
  increases as well while their average mass decreases. This stems
  from the fact that high-Mach number flows produce many and strong
  shocks on intermediate to small spatial scales, leading to a
  highly-fragmented density structure. We also show that the CMD from
  purely turbulent fragmentation does not follow a single power-law,
  but it may be described by a function that changes continuously its
  shape, probably more similar to a log-normal function.  The CMD in
  supersonic turbulent flows does not have a universal slope, and
  as consequence, cast some doubt on attempts to directly relate the CMD
  to a universal Initial Mass Function.

\end{abstract}


\section{Introduction}\label{intro:sec}

An isothermal supersonic shock with a Mach number $\Ms$ creates
density enhancements of $\rho_1/\rho_0 = \Ms^2$, where $\rho_1$ and
$\rho_0$ are the densities of the post- and pre-shocked gas (e.g.,
Spitzer 1978).  Since molecular clouds are turbulent and supersonic,
it can be expected that their internal density structure is, at first
order (i.e., neglecting gravitational or thermal fragmentation), a
direct consequence of the fragmentation by the chaotic, supersonic
velocity field (see, e.g., the reviews by V\'azquez-Semadeni et
al. 2000; Mac Low \& Klessen 2004; Scalo and Elmegreen 2004 and
references therein), a process which has been called {\it turbulent
fragmentation}.  Thus, it is reasonable to expect that supersonic
turbulence plays a crucial role in determining the mass distribution
of dense cores.  In fact, the gravoturbulent scenario of star
formation suggests that the cores are formed by compressible turbulent
motions inside molecular clouds and that some of those cores may
become gravitationally unstable and form stars, while others will
redisperse in the ambient medium (Sasao 1973; Hunter
\& Fleck 1982; Elmegreen 1993; Ballesteros-Paredes, V\'azquez-Semadeni
\& Scalo 1999; Klessen et. al. 2000; Padoan et al. 2001; Padoan \&
Nordlund 2002, hereafter PN02). 

On the other hand, the mass distribution of young stars follows a
well-known distribution called the Initial Mass Function (IMF).  For
stellar masses $M\ge 1\,M_{\odot}$ it shows a power-law behavior
$dN/d{\log}M \propto M^\Gamma$, with slope $\Gamma = -1.3$ (Salpeter
1955; Scalo 1998; Kroupa 2002; Chabrier 2003).  Understanding the
origin of the IMF is one of the fundamental goals of a complete theory
of star formation.  Although important progresses have been achieved
on the observational determination of the IMF, there are still several
proposed models to explain it (see reviews by Meyer et al. 2000; Mac
Low \& Klessen 2004, and references therein), and there is no
agreement in the community for a standard one.  One of the more recent
models suggests that the IMF properties are a direct consequence of
the core mass distribution (CMD).  Observational works (e.g., Motte et
al. 1998; Testi \& Sargent 1998) have reported a slope of the
high-mass wing of the dense core mass distribution (or mass spectrum)
that is similar to the slope of Salpeter's IMF, suggesting that those
cores are the direct progenitors of single stars.  Since stars are
born from dense cores this idea is, in principle, tempting.  However,
there is a large number of physical processes that may play an
important role during the core fragmentation and the protostellar
collapse (see., e.g., Klessen \& Burkert 2000; Goodwin, Whitworth \&
Ward-Thompson 2004; Bate \& Bonnell 2004). These make it unclear
whether a single core will give birth to one or more stars, and what
determines the masses of individual stars within a single core.  Some
of these processes are: (a) The mass distribution of cores changes
with time as cores merge with each other (e.g., Klessen 2001; Schmeja
\& Klessen 2004). (b) Cores generally produce not a single star but
clusters of stars, and so the relation between the masses of cores and
those of individual stars is unclear (e.g., Larson 1985, Hartmann
2001, Goodwin et al.\ 2004). In addition, there may be (c) competitive
accretion influencing the mass-growth history of individual stars
(see, e.g., Bate \& Bonnell 2004), (d) stellar feedback through winds
and outflows, or (e) changes in the equation of state introducing
preferred mass scales (e.g., Scalo et al.\ 1998; Li, Klessen \& Mac
Low 2003; Jappsen et al.\ 2005; Larson 2005).

Besides the uncertainties mentioned above, there are other important
caveats when looking for a direct relationship between the CMD and the
IMF.  For instance, even though some observational and theoretical
works for dense, compact cores fit power-laws in the high-mass wing of
the CMD, the actual shape of those CMDs is not necessarily a
single-slope power-law, but a function whose slope varies in a more
continuous way, frequently similar to a log-normal distribution.  From
a theoretical point of view, PN02 have argued that the mass
distribution of dense cores generated by turbulent fragmentation
follows closely the Salpeter distribution of intermediate- to
high-mass newborn stars, with a slope of $\sim -1.3$ and that the
slope of the CMD depends only on the slope of the turbulent energy
spectrum.  These results have been taken as a proof that turbulent
fragmentation is essential to the origin of the stellar IMF.  Recent
numerical studies by Tilley \& Pudritz (2004) and Li et al.\ (2004)
have reported that the prediction of the PN02 model for the CMD agrees
with the results of their simulations.  We consider that some
cautionary remarks concerning these theoretical results are necessary.
First, the dynamical range of the CMD where a power-law with slope
$-1.3$ is appropriate is usually smaller than one order of magnitude.
Second, the reported CMDs are calculated at one single epoch, instead
of being averaged over several timesteps.  Due to the fact that
stochastic fluctuations might be non-negligible in a single frame (see
\S\ref{convergence_considerations:sec}), determining CMDs at one
single time is prone to significant statistical fluctuations.  This is
particularly relevant when the core statistics is small, as in Li et
al.\ (2004).  Third, the behavior of the CMDs presented in those works
seems to be a continuous change in the slope of the CMD, from zero at
the maximum of the histogram, to large negative slopes in the
high-mass range.  In fact, Gammie et al. (2003) mentioned that the
high-mass wing of the core mass spectrum in their simulations has a
slope that appears to be consistent with the Salpeter law, although
other non-power-law forms for the mass spectrum may well be
consistent.  Finally, in these works there is no systematic study of
the dependence of the CMDs with the Mach number.  For example, Tilley
\& Pudritz (2004) present simulations with Mach numbers of 2 and 5,
and indeed a close inspection of their Fig.~17 shows that the CMDs are
different in both models, even though they can be fitted with the
power-law with the slope of $-1.3$ in a small dynamical range.
%RSK
%Most power-law fits to the slope of the CMD reported in the literature
%appear appropriate only for a small range of masses: At the peak of
%the CMD the slope certainly is zero and then decreases continuously as
%the mass increases. In almost all cases there is a part which has a
%value $\Gamma = -1.3$ similar to the stellar IMF.

As has been discussed by Klessen (2001) and by Schmeja \& Klessen
(2004), at a given Mach number the shape of the mass spectrum of dense
cores changes with the wavenumber at which turbulence is driven.  In a
similar way, the slope of the mass spectrum may very well vary with
the root-mean-square (RMS) Mach number $\Ms$ of the flow, which is
related to the kinetic energy density of the system.  In fact, using
numerical simulations, Kim \& Ryu (2005, see also Cho \& Lazarian
2004) had found that the density power spectra of isothermal,
turbulent flows, vary with the Mach number, becoming shallower when
the Mach number is increased.  This flattening is the consequence of
the dominant density structures of filaments and sheets.  This occurs
because, in a supersonic turbulent flow, density fluctuations are
built up through a sequence of jumps, produced by a succession of new
shocks within previously shocked regions (V\'azquez-Semadeni 1994;
Passot \& V\'azquez-Semadeni 1998).  Thus, when increasing the Mach
number, the time frequency and strength of shocks inducing density
enhancements also increase.  Since the density power spectra becomes
shallower (Kim \& Ryu 2005), this means that the flow contains smaller
and denser structures.  Thus, it can be expected that the CMD depends
on the Mach number.

In the present work we test this idea by using two different numerical
schemes that model the interior of molecular clouds.  In our
simulations we vary the value of $\Ms$ from 1 (trans-sonic turbulence)
to 15 (supersonic, highly compressible turbulence), and show that the
form of the CMD does indeed depend on the Mach number.  The plan of
the paper is the following: in \S\ref{simulations:sec} we briefly
introduce the simulations used in the present work, and discuss the
method for finding clumps, showing that turbulent systems reach
statistical equilibrium after one turbulent crossing time.  We also
discuss, the importance of averaging the CMD over several timesteps in
order to erase statistical fluctuations.  In \S\ref{results:sec} we
present our results.  \S\ref{discussion:sec} discusses the application
of the PN02 model to isothermal hydrodynamic turbulent fields and the
relationship between the core mass distribution and the IMF. Finally,
in \S\ref{conclusions:sec}, we summarize our results.

\section{Simulations}\label{simulations:sec}

In order to study how the mass spectra of the cores depend on the Mach
number, we analyzed two different sets of simulations made with two
distinct numerical schemes.  The first set uses a Total Variation
Diminishing (TVD) method, and the second one a Smoothed Particle
Hydrodynamics (SPH) approach.  All the simulations presented here are
three-dimensional and isothermal.  We concentrate on small subregions
within a much larger cloud. Therefore, periodic boundary conditions
are adopted in both numerical schemes.  Since we are interested in the
fragmentation produced by the sole presence of turbulence, the
simulations do not include self-gravity and magnetic fields.  Being
isothermal, the simulations are scale-free, and the renormalization is
somehow arbitrary, but in principle they reproduce the behavior of
molecular clouds of sizes ranging from less than 0.1 pc up to
10--20~pc.  Table~\ref{tabla:models} presents the properties of each
run.  In the first four columns we list the name of the run, its RMS
Mach number (average over several timesteps), the numerical method
used, and the resolution.  The remaining columns will be introduced in
\S\ref{convergence_considerations:sec}.

\begin{deluxetable}{c c c c c c c c c c}
 \tablecolumns{10}
 \tabletypesize{\scriptsize}
 \tablecaption{Model properties. \label{tabla:models}}
 \tablewidth{0pt}
 \tablehead{
      \colhead{Name}  &
      \colhead{RMS Mach Number}  &
     \colhead{Method}    &
     \colhead{Resolution\ala a} &
%     \colhead{Mass\ala b} &
     \colhead{\# frames} & 
     \colhead{$t_i$\ala b} &
     \colhead{$t_f$\ala b} &
     \colhead{$\Delta t_{\rm frame}$\ala c} &
%     \colhead{$\Delta t_s$\ala d}
%\\
\\
 }
 \startdata
 TVD1          &  0.97  & TVD &$256^3$    & 8  & 1.2   & 5     & 0.4    \\ % & 0.4  \\
 TVD3          &  2.87  & TVD &$256^3$    & 12 & 1.2   & 4.5   & 0.3    \\ % & 0.9  \\
 TVD6          &  6.2   & TVD &$256^3$    & 18 & 1.2   & 6     & 0.3    \\ % & 1.8  \\
\\
 SPH3          &  3.4   & SPH &205,000    & 96 & 1.02  & 4.25  & 0.034  \\ % & 0.102\\
 SPH6          &  5.7   & SPH &205,000    & 64 & 1.008 & 4.536 & 0.056  \\ % &0.336\\
 SPH9          &  9.1   & SPH &205,000    & 77 & 3.375 & 6.795 & 0.0405 \\ % &0.405\\
 SPH15         &  14.98 & SPH &205,000    & 50 & 1.05  & 4.725 & 0.075  \\ % &1.125\\
\\
 TVD-HR        &  3     & TVD &$512^3$    & 12 & 1.2   & 4.5   & 0.3    \\ % & 0.9 \\
 SPH-HR\ala e  &  3.8   & SPH &9,938,375 & 6  & 1.368 & 1.558 & 0.038  \\ % &XX   \\
 \enddata
 %%JK
 \tablenotetext{a}{Numbers of cells for the TVD scheme, and particles
 for the SPH scheme.} 
 \tablenotetext{b}{In units of the turbulent crossing time.}
% \tablenotetext{c}{ime spacing between analyzed frames, in sound
% crossing timesteps.}
 \tablenotetext{d}{Time spacing between frames, in turbulent
 crossing timesteps.}
\tablenotetext{e}{Run performed with Gadget.}
% %\tablecomments{\\ tablecomments here, if there are.}
 \end{deluxetable}

\subsection{Total Variation Diminishing}

The TVD scheme is a second-order accurate upwind scheme whose
implementation for isothermal flows is described in detail by Kim et
al.\ (1999).  Here we just mention that the simulations performed used
a turbulence random driver following the method presented in Stone,
Ostriker \& Gammie (1998).  Each driving velocity component is
generated in Fourier space, and its power spectrum has a peak at a
large scale (specifically at wavenumber $\kfor=2\,(2\pi/L_0)$, where
$L_0$ is the one-dimensional size of the computational box).  The
kinetic energy input rate is adjusted in order to maintain a roughly
constant RMS sonic Mach number.  The three simulations used in the
present work have a resolution of 256\ala 3 cells, and their
three-dimensional RMS Mach numbers are 1, 3 and 6. An additional
high-resolution simulation (512\ala 3 cells) at Mach 3 has been
performed, in order to check that the log-normal type shape of the CMD
is not an artificial result of the low-resolution simulation.

\subsection{Smooth Particle Hydrodynamics}

The SPH method is a Lagrangian scheme in which the fluid is
represented by an ensemble of particles, and flow quantities are
obtained by averaging over an appropriate subset of the SPH particles
(Benz 1990). The method is able to resolve large density contrasts as
particles are free to move, and so naturally the particle
concentration increases in high-density regions.  Thus, this method
has spatially varying spatial resolution as a function of density (see
Ossenkopf, Klessen \& Heitsch 2001).  Details on the method can be
found in Klessen (2001; see also Klessen, Heitsch, \& Mac~Low 2000;
and Klessen \& Burkert 2000).  In the four SPH simulations analyzed in
this paper the three-dimensional RMS Mach numbers are 3.4, 5.7, 9 and
15.
%RSK: I think we have too many footnotes. I thus merged with the
%previous two...
%
%\footnote{Note that if we use the spherical definition of
%mass, $M_J = 4 \pi /3 \rho (\lambda/2)^3$, as in Klessen (2001), the
%total mass in the computational box is 120 Jeans masses}.  
Again, we drive turbulence at large scales, but in this case in a
narrow range of wavenumbers $\kfor\le2\,(2\pi/L_0)$.

As in the case of the TVD method, we performed an additional high
resolution SPH simulation (9,938,375 particles) at Mach 3.8, in order
to check that the shape of the CMD is not an artificial result of the
resolution. In this case we used the parallel code GADGET (Springel,
Yoshida \& White 2001), in which the same turbulent driving scheme as
in Klessen (2001) has been implemented (see Jappsen et al. 2005 for
details).

\subsection{Clumpfinding particularities}\label{analysis:sec}

We analyze all the simulations over many equally spaced timesteps
covering several turbulent crossing times, $\tturb$ (where $\tturb$ is
defined as by the one-dimensional size of our computational box
$L_{\rm box}$ divided by the RMS velocity of the field --i.e., the
Mach number).  To compute the CMD at each timestep, we use an
adapted version of the Williams et al. (1994) clumpfinding algorithm
to find cores in the three-dimensional density data-cubes calculated
with the TVD scheme.  This modified version allows us to find clumps
even in large ($512^3$) density cubes.  For the particle data, we used
its equivalent clumpfinding algorithm, developed by Klessen \& Burkert
(2000, see their Appendix A).  In both schemes, we used logarithmic
contours separated by half order of magnitude.  The parameters of the
clumpfinding algorithms in each data set are somehow different: for
the TVD runs, the lower density threshold is 2 times the mean density,
and for the SPH runs the lowest density contour is 10\alamenos{2.5}
($\sim$~1/50 the mean density).  In both cases, the separation of
contours is logarithmic, but while in the TVD runs the increase of
mass is in steps of half order of magnitude, in the SPH we adopted the
same procedure as Klessen (2001), having only 10 logarithmic
isocontours starting from the lowest contour and increasing the
exponent by 0.5.  Although the details of the CMD obtained with
different parameters of the clumpfinding algorithms may change (e.g.,
exact mass of each core),
%the main result obtained here, that the CMD
%from pure turbulence varies with the Mach number, 
the trend of the CMD variation with Mach number does not depend on the
numerical method and/or the details of the clumpfinding scheme (see
\S\ref{results:sec}). As the main purpose of this work is to observe
this trend, the conclusions we reach are also independent of those
choices.

%\subsection{Analysis}
%RSK modified
\subsection{Temporal convergence
considerations}\label{convergence_considerations:sec}
\begin{figure}
%\plotone{all_vs_time.ps}
\plotone{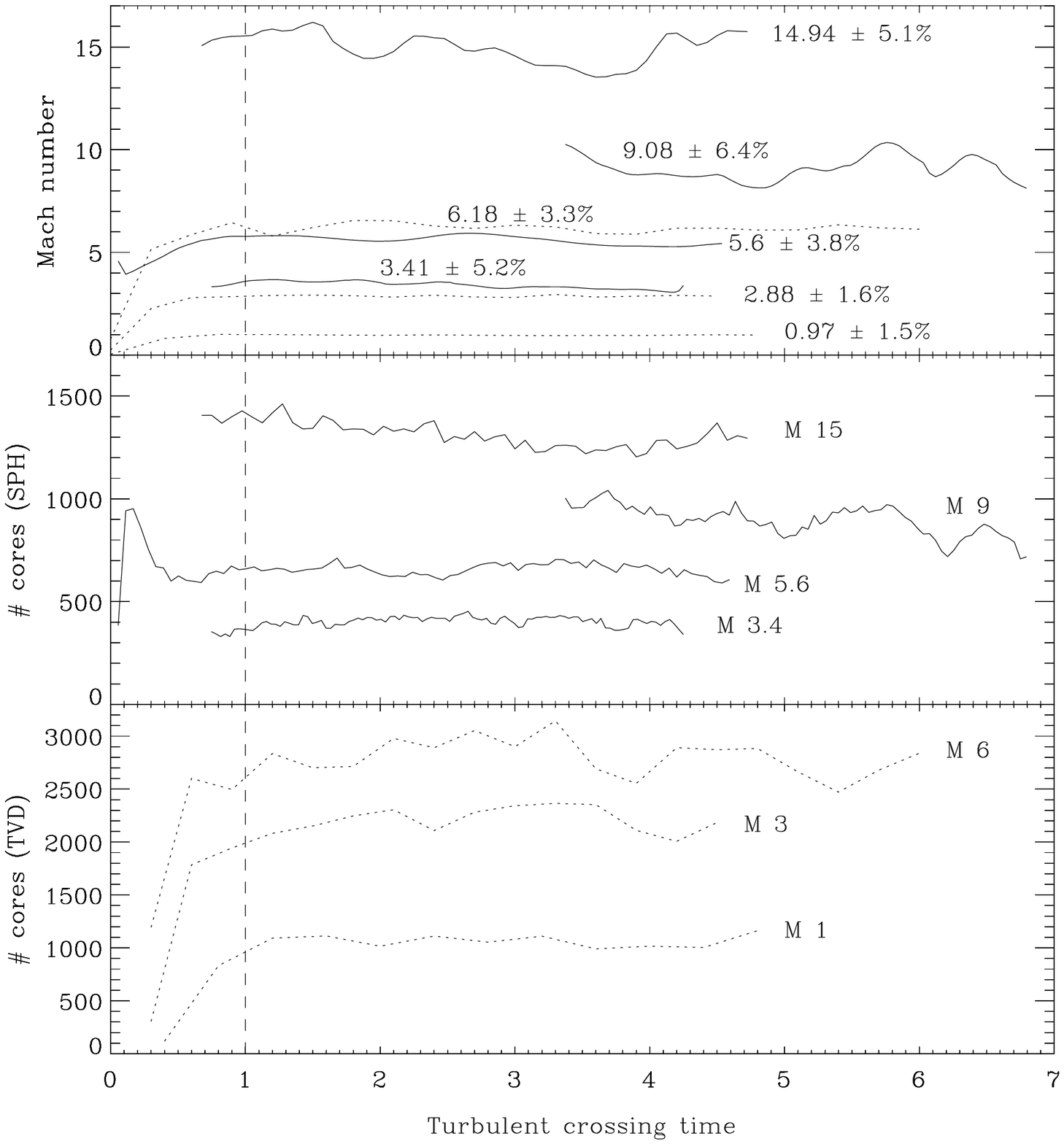}
 \caption{Temporal evolution of the Mach number (upper panel), number
 of cores found in the SPH models (middle panel) and in the TVD models
 (lower panel).  Solid lines denote SPH runs, and dotted lines denote
 TVD runs. Quoted numbers in the upper panel give the mean Mach
 number, averaged over the frames whose times are larger than 1
 turbulent crossing time.  (see Table~\ref{tabla:models}).  Note that
 after 1 turbulent crossing time, the fluid has reached equilibrium,
 since (a) the Mach number becomes nearly constant, with fluctuations
 not larger than 6.5\%\ (upper panel), and the number of cores has
 reached its typical value. Note that still fluctuations up to 20\%\
 in the number of cores may be found, showing the importance of
 averaging over several timesteps when calculating CMDs.
\label{all_vs_time:fig}}
\end{figure}

Since the simulations start from arbitrary random initial conditions
with density and velocity physically decorrelated, it is convenient,
first, to investigate the time at which the fluid reaches statistical
equilibrium.  We have studied the temporal evolution of the energy
spectrum (not shown here for reasons of space), and found that after
one turbulent crossing time, the energy spectrum of each run does not
vary substantially.  To verify that, in addition to the energy
spectra, at this time the simulations have reached statistical
equilibrium, we show in Fig.~\ref{all_vs_time:fig} the Mach number as
a function of time ({\it upper panel}), as well as the number of cores
in the SPH ({\it middle panel}) and TVD ({\it lower panel})
simulations.  From this figure, where solid lines represent the SPH
data, and dotted lines represent the TVD data, it becomes clear that,
in fact, after 1 $\tturb$, the RMS Mach number (upper panel) in
the flow approaches to a constant value\footnote{Note however, that
$\Ms$ still shows variations of order 5\% around the mean value,
specially in the SPH data.  This is due to compressibility of the
medium and the adopted fixed large-scale driving pattern, as discussed
by Klessen \& Lin (2003). The deviations from the mean value between
individual time frames decrease with decreasing RMS Mach number.}. The
quoted numbers give the RMS Mach number $\Ms$, averaged over all
frames for which $t \ge 1$ $\tturb$.  Similarly, in the middle and
lower panels, where the number of cores is plotted as a function of
time, it is again seen that the transients are given only for the
first turbulent crossing time.  After this time, each frame may give
statistical fluctuations of the order of 20\%\ in the number of cores.

An important observation from the middle and lower panels in
Fig.~\ref{all_vs_time:fig} is that, given the fact that important
fluctuations in the number of cores are present, making conclusions
based on a CMD computed from individual frames is risky.  To prevent
statistical bias when reporting the shape of the CMD, it is therefore
important to calculate it as an average over at least a few turbulent
crossing times.  It is convenient to note that, if the average were
calculated over a fraction of turbulent time, the statistical
fluctuation of the small cores will be erased, but the statistical
bias of the larger cores will not, since in a fraction of one
turbulent time the small cores may change considerably, but the larger
ones not.

Getting back to our Table \ref{tabla:models}, in its last four columns
we show, for each simulation, the number of frames used in the present
analysis (column 5), the initial and final times (in units of
$\tturb$) used in the analysis (columns 6 and 7), and the time
separation between frames, in units of turbulent crossing times
(column 8).

\section{Results}\label{results:sec}
 
\begin{figure}
%\plotone{f1.eps}
\plotone{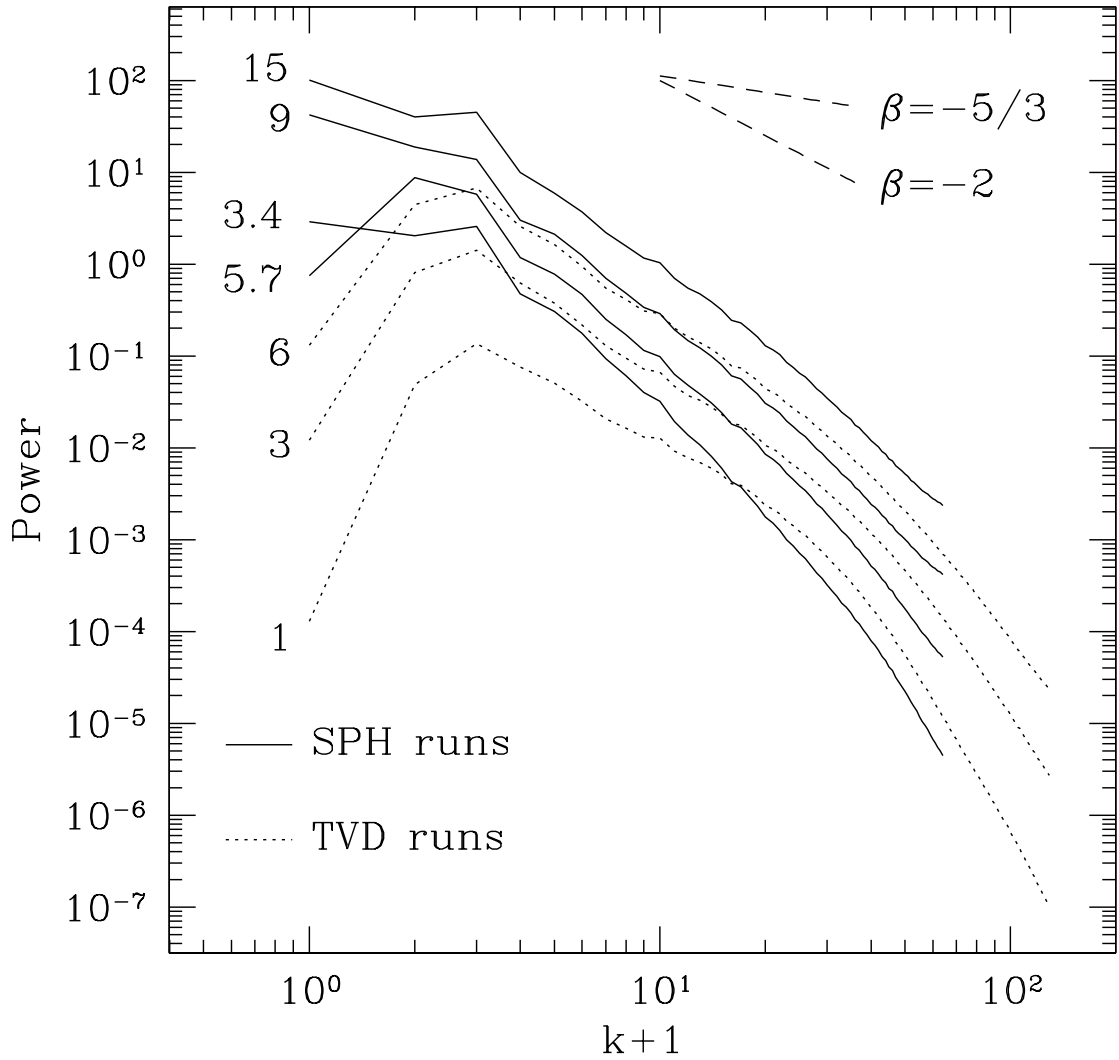}
 \caption{Energy spectrum for each simulation. The {\it solid lines}
 correspond to SPH simulations and the {\it dotted lines} to TVD
 simulations. The {\it dashed lines} have a slope of $-2$ and $-5/3$.
\label{energy:fig}}
\end{figure}

Figure \ref{energy:fig} shows the time-averaged energy spectrum of
each simulation.  For comparison, the expected slopes for an
incompressible flow ($\beta = -5/3$) and for a shock-dominated one
($\beta = -2$) are indicated by dashed straight lines.  
%In this
%figure, it can be seen that the run with $\Ms=1$ has, in fact, an
%energy spectrum with slope close to $-5/3$ (incompressible), while the
%remaining simulations have steeper energy spectra with slopes more
%similar to $-2$. 
The kinetic energy density is proportional to the total area below the
curve. This area grows as the $\Ms$ increases.  Note that the energy
spectra for the SPH runs have somewhat shallower slopes than those for
the TVD runs.  However, this can be understood as a consequence of the
considerable differences between the two numerical schemes.  The
former is a particle-based method while the latter uses finite
differences on a equally-spaced mesh.  Both methods are complementary
(see also the discussion in Klessen et al.\ 2000), and we base our
discussion solely on trends that are common in both sets of
simulations.  The fact that both numerical schemes, TVD and SPH, give
rise to the same behavior of the CMD with RMS Mach number (see
\S\ref{results:sec}), reassures us to consider the observed effect of
varying $\Ms$ as a real physical trend.

%\section{Relation between Density Structure and Mach Number}\label{results:sec}

\begin{figure}
%\plottwo{f2a.ps}{f2b.ps}
\plottwo{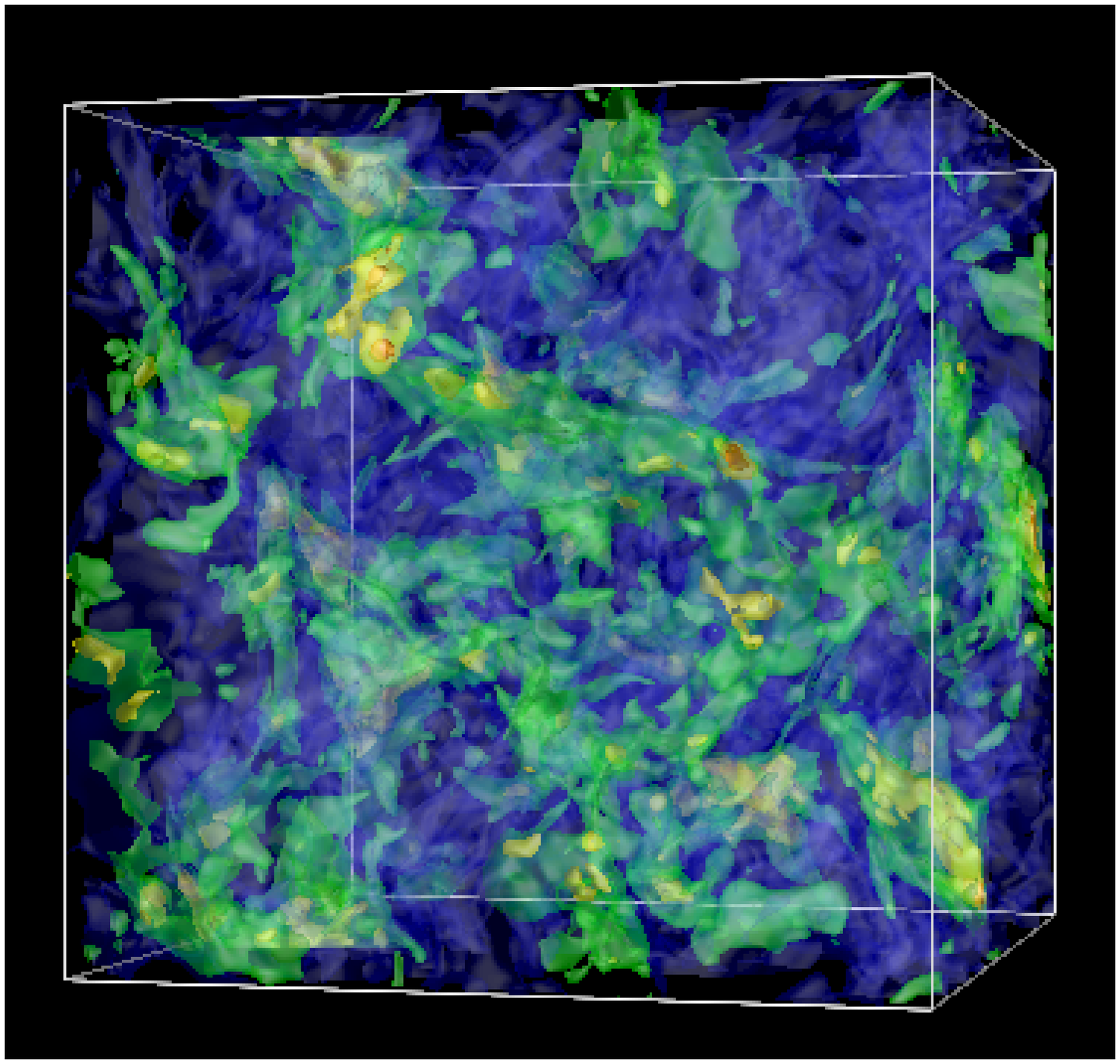}{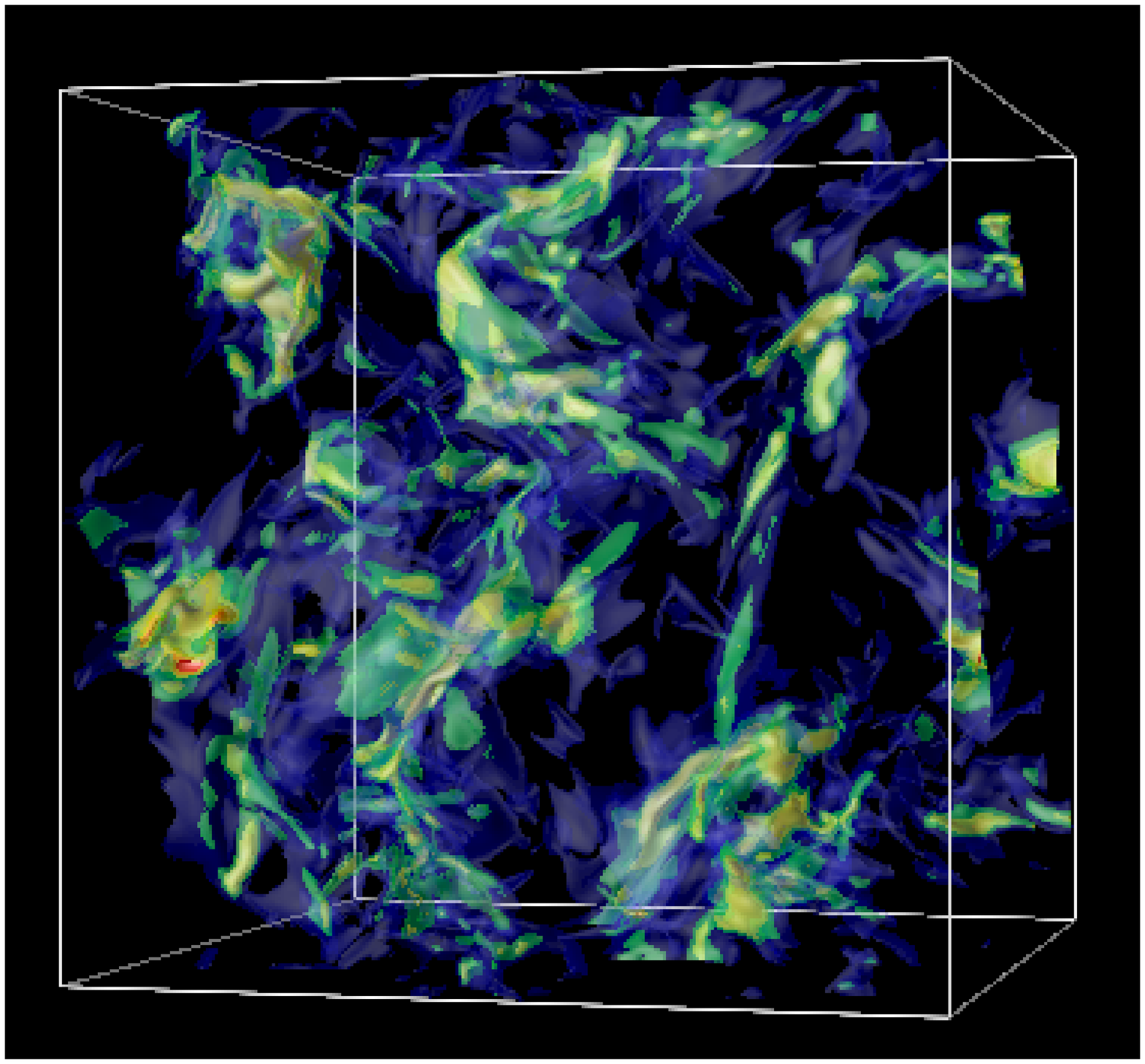}
 \caption{Snapshots of the TVD runs with (a) $\Ms=1$ at
 $t=5\,\tau_{\rm turb}$, (b) $\Ms=6$ at $t=4.5\,\tau_{\rm turb}$.
 Note how the turbulent fragmentation of the medium is different for
 different Mach numbers.  Blue, green, yellow, orange, and red colors
 represent isodensity surfaces with 1, 2, 3, 4 and 5 times the mean
 density in frame (a). In frame (b), the same order of colors
 represents isodensity surfaces with 5, 11.25, 17.5, 23.75 and 30
 times the mean value. 
\label{tvd:cubes:fig}}
\end{figure}

%\subsection{The shape of the core mass distribution changes with the Mach 
%number}
%
%RSK modified
Figures \ref{tvd:cubes:fig}a, b show snapshots of the density field in
two TVD simulations with $\Ms=$1 and 6, respectively.  The snapshots
have been acquired at $t=5\,\tau_{\rm turb}$ ($\Ms=1$) and
$t=4.5\,\tau_{\rm turb}$ ($\Ms=6$).
%RSK this is also said in the caption, it adds nothing here in the
%text....
% The
% left panel shows isodensity surfaces with five different values from
% $n=n_0$ to $n=5n_0$ with a step of $n_0$.  Those values are mapped
% with blue, green, yellow, orange, and red colors, respectively. For
% Fig.~\ref{tvd:cubes:fig}b the same order of colors represents the
% density surfaces with 5, 11.25, 17.5, 23.75, and 30 times $n_0$.  
It is evident that the medium is more strongly fragmented in the high
Mach number case. A similar conclusion holds for the SPH data cubes.
The interactions of supersonic shocks and shocklets in high-RMS Mach
flows produce strongly localized density enhancements, and
consequently give rise to high degrees of fragmentation {\it even} on
scales much smaller than the turbulent driving scale.

\begin{figure}
\plotone{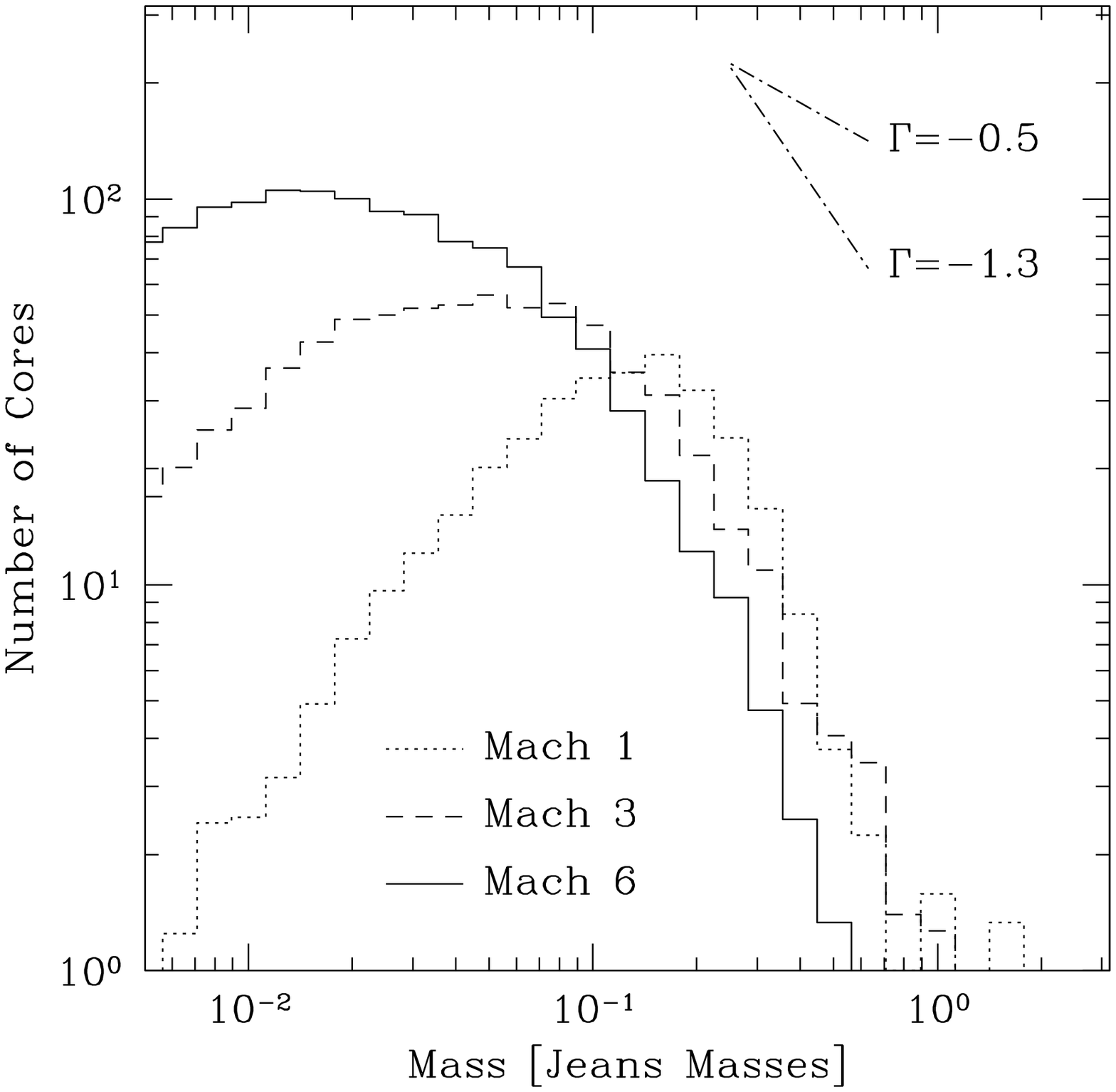}
%\plotone{f3.eps}
% \plotone{cmd_clf_jsk_ro_2p02.ps}
 \caption{Time averaged core mass distribution resulting from TVD
 simulations.  The two {\it dot-dashed lines} with slopes of $-0.5$
 and $-1.3$ are shown for the comparison purpose with the mass
 functions of Giant Molecular Clouds and protostellar cores,
 respectively. \label{tvd:histo:fig}}
\end{figure}

\begin{figure}
%\plotone{f4.eps} 
% \plotone{histo_sph.eps}
 \plotone{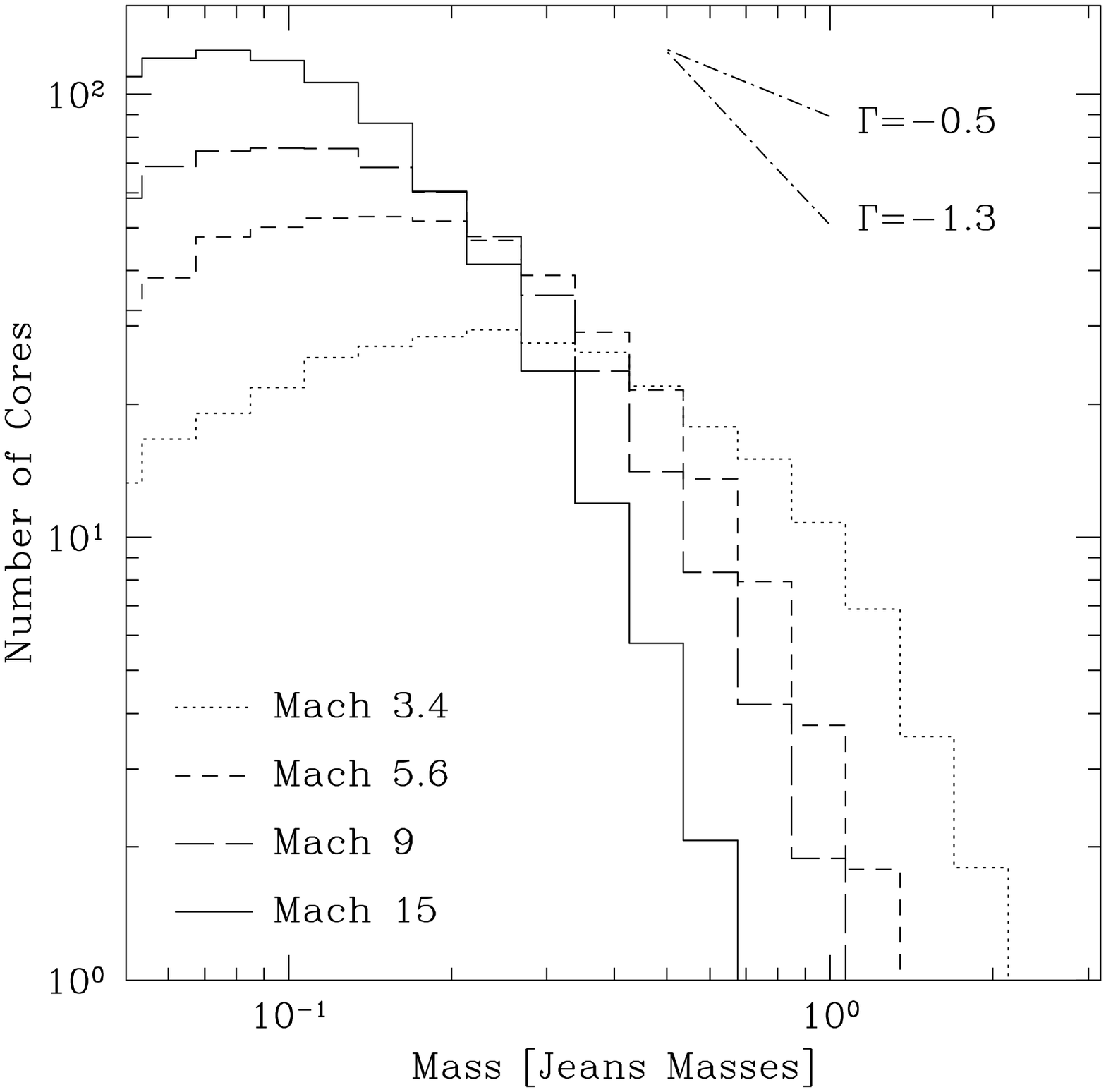}
 \caption{Time averaged core mass distribution resulting from SPH
 simulations.  The two {\it dot-dashed lines} with slopes of $-0.5$
 and $-1.3$ are shown for the comparison purpose with the mass
 functions of Giant Molecular Clouds and protostellar cores,
 respectively.
\label{sph:histo:fig}}
\end{figure}

\begin{figure}
%\plottwo{f5a.eps}{f5b.eps}
\plottwo{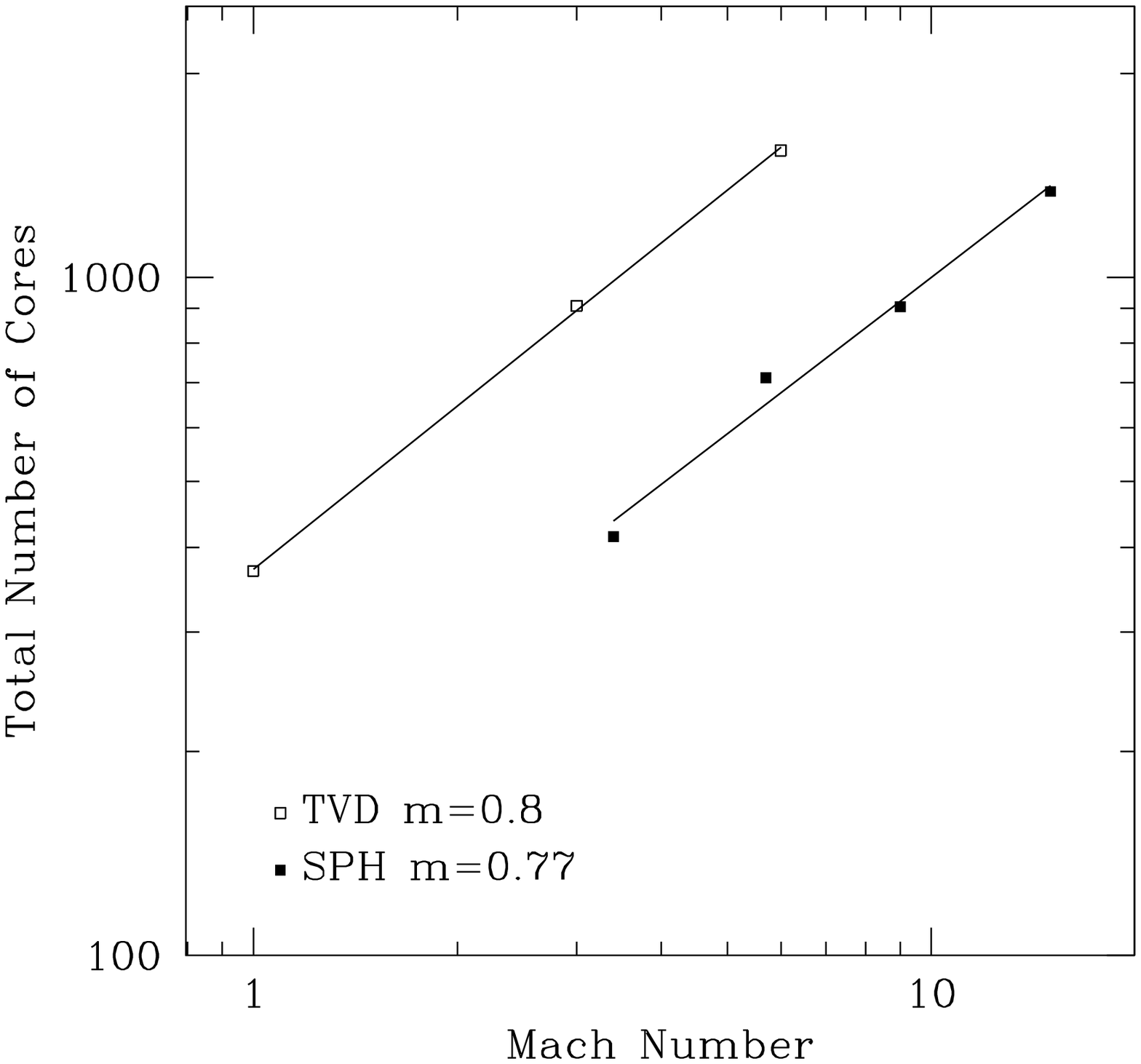}{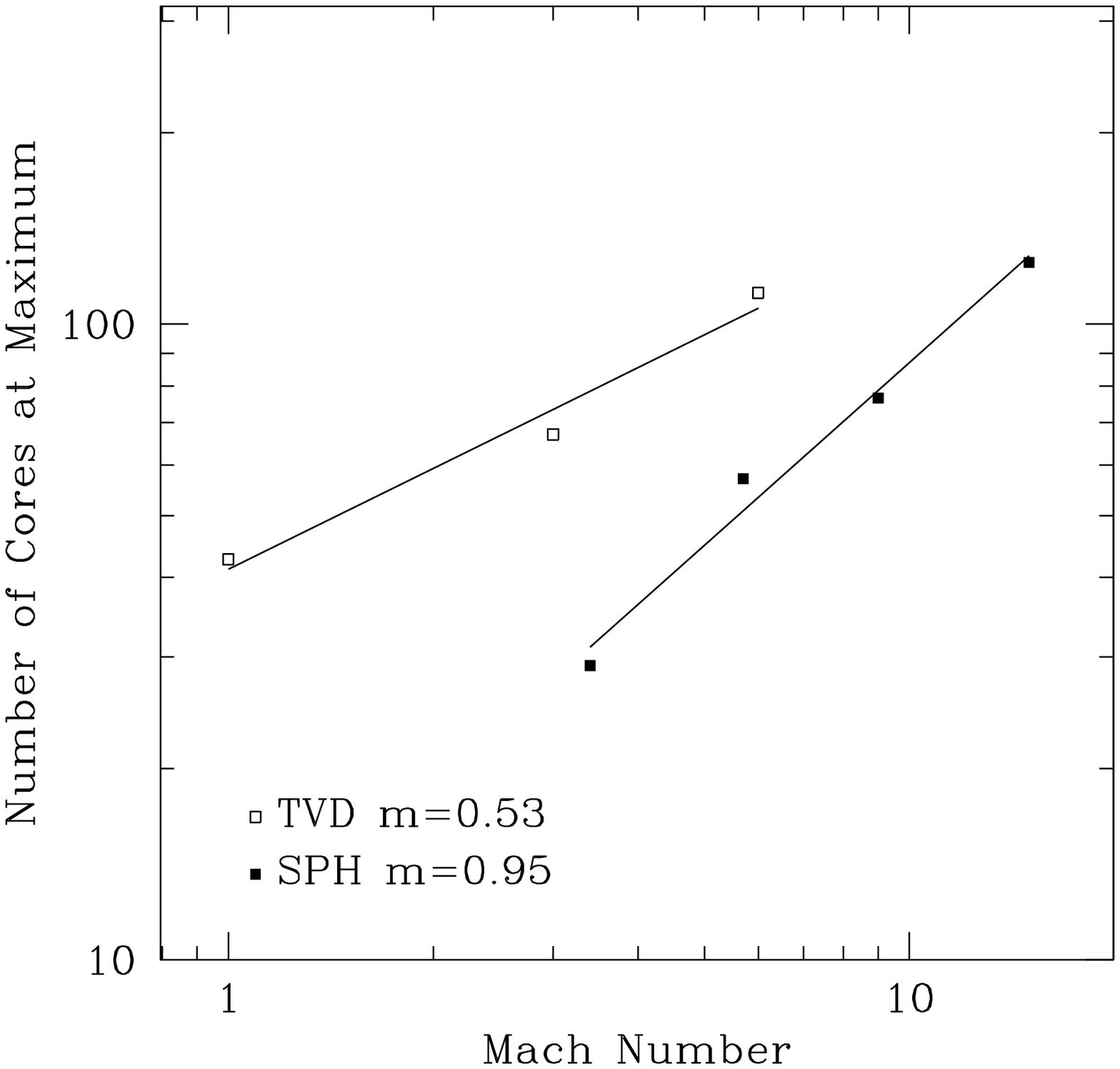}
 \caption{(a) Total number of cores, and (b) Number of cores at
 maximum of the distribution, as a function of the mach number $\Ms$.
 Filled (open) squares denote results from SPH (TVD) runs. Solid lines
 represent least-square fits to the data.  $m$ represents the slope of
 the fit.  Note that, for each dataset, the number increases as $\Ms$
 increases.
\label{number_vs_mach:fig}}
\end{figure}

\begin{figure}
%\plottwo{f6a.eps}{f6b.eps}
\plottwo{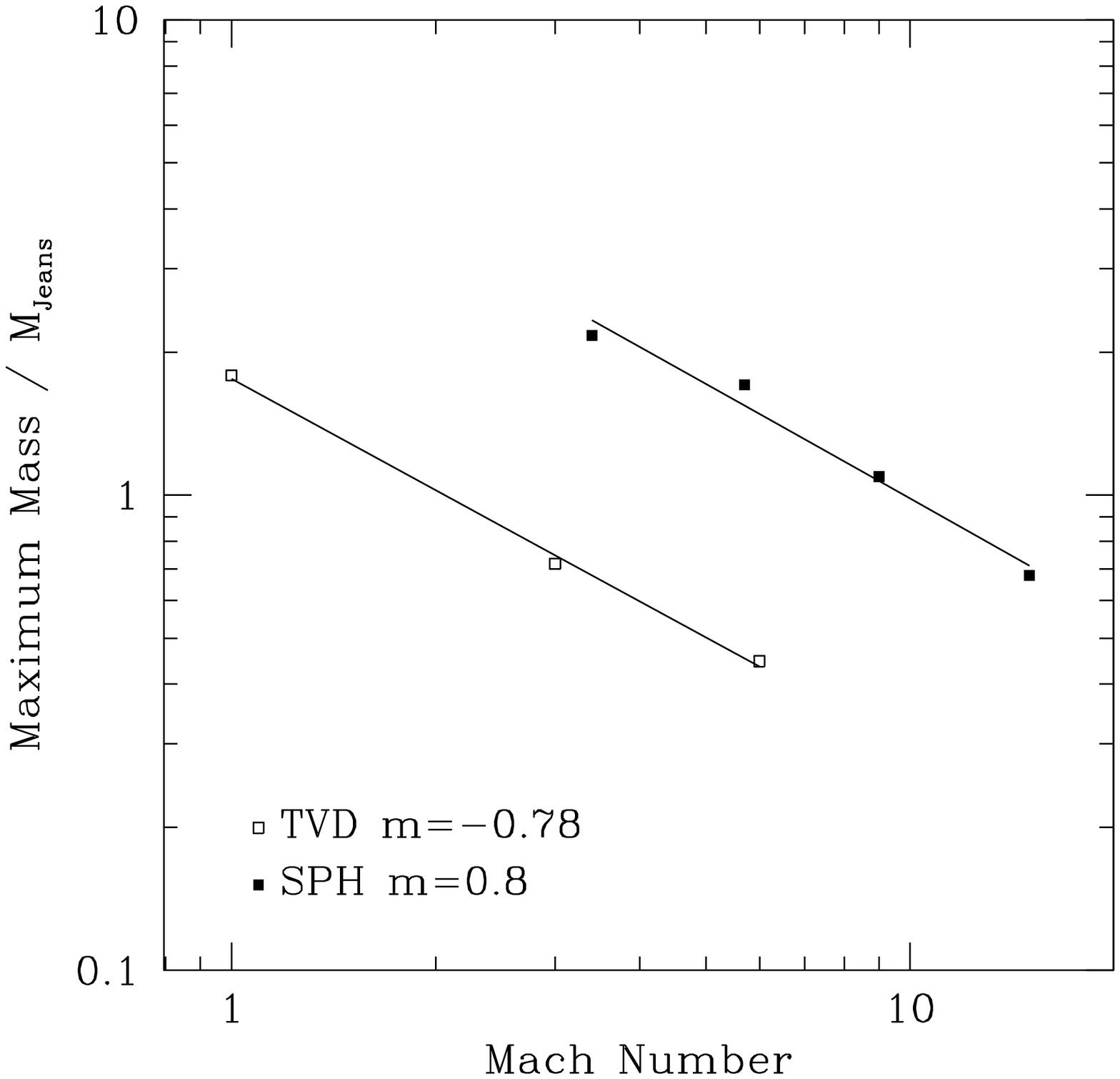}{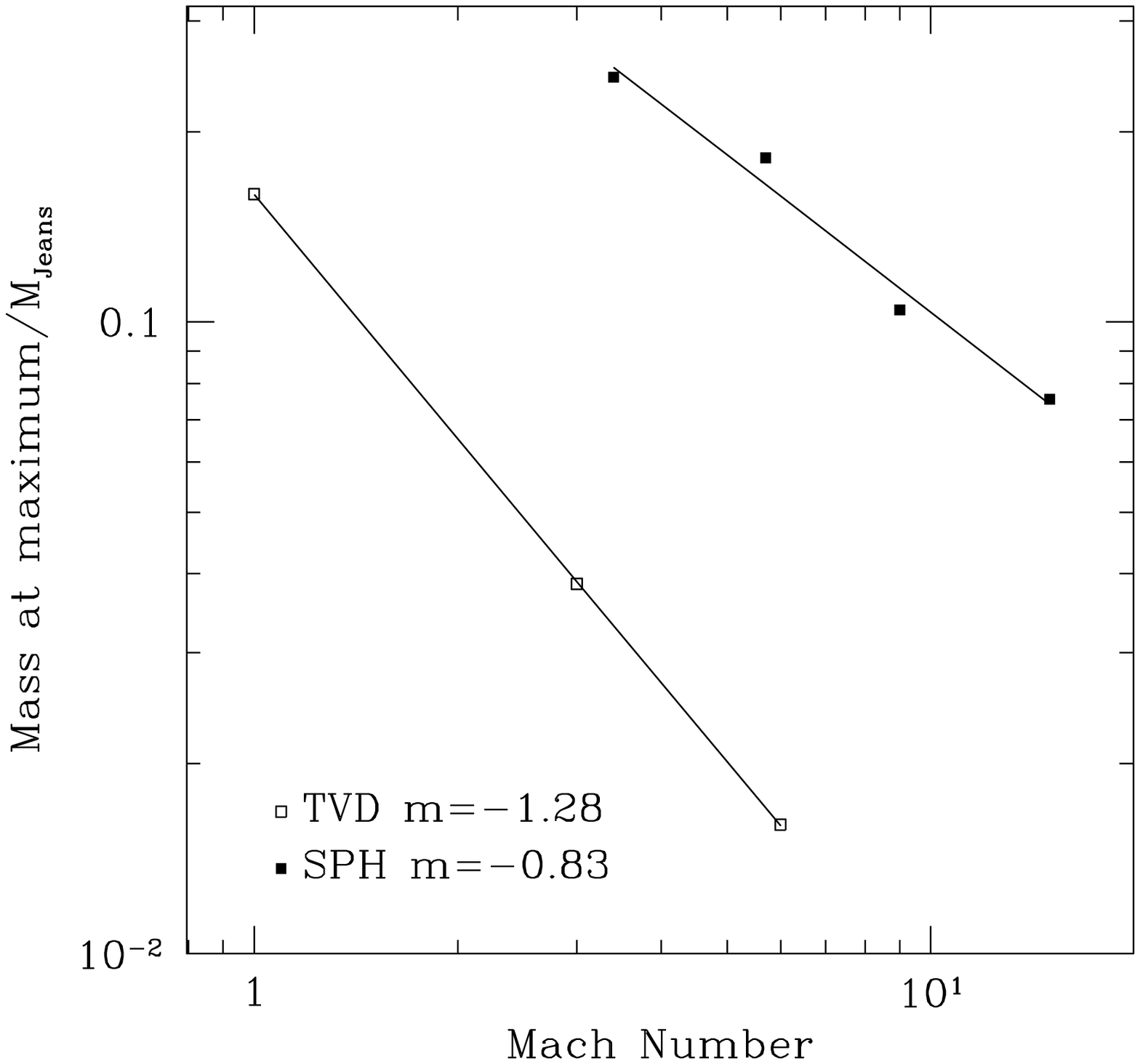}
 \caption{(a) Mass of the most massive core, and (b) mass at which the
 maximum of the histogram occurs as a function of $\Ms$.  Filled
 (open) squares denote results from SPH (TVD) runs. Solid lines
 represent least-square fits to the data.  $m$ represents the slope of
 the fit.  Note that, for each numeric scheme, the mass decreases as
 $\Ms$ number increases.
\label{mass_vs_mach:fig}}
\end{figure}

%RSK modified...
To quantify that further, we present in Figs.~\ref{tvd:histo:fig} and
\ref{sph:histo:fig} the resulting core mass spectra resulting from
the TVD and the SPH models, respectively.  The total mass in the
computational box for all the models has been renormalized to 64 Jeans
masses\footnote{Even though we do not take the self-gravity of the gas
into account, we give the total mass in units of the Jeans mass.
Other choices of the mass unit are possible and fully equivalent. The
reason for quoting the mass in Jeans masses is to allow for direct
comparison with recent numerical works, e.g., by Klessen (2001),
Gammie et al. 2003, Tilley \& Pudritz (2004); Li et al. (2004), as
well as with a future contribution of the analysis in the
self-gravitating case.  Note that we adopt a cubic definition of the
Jeans mass, $M_J = \rho \lambda^3$. This differs from the spherical
definition, $M_J = 4 \pi /3 \rho (\lambda/2)^3$, by a factor
$\sim2$.}. The CMD clearly varies with $\Ms$.  In particular, the
total number of cores as well as the peak value of the histogram
increase with $\Ms$.  To illustrate this result more clearly we show
in Fig.~\ref{number_vs_mach:fig} (a) the total number of cores and,
(b) the maximum value reached by the histograms as function of $\Ms$.
In this figure, filled and open squares correspond to the SPH and the
TVD models, respectively.  Solid lines in each panel are least-square
fits.  As a consequence of mass conservation (recall that each TVD and
SPH simulation has the same total mass) the typical core mass shrinks
as the number of cores increases.  This result is shown in
Fig.~\ref{mass_vs_mach:fig}, where we plot (a) the mass of the most
massive core and, (b) the mass at which the histogram peaks, both as
function of $\Ms$.

\begin{figure}
%\plotone{histogramas_HR.ps}
\plotone{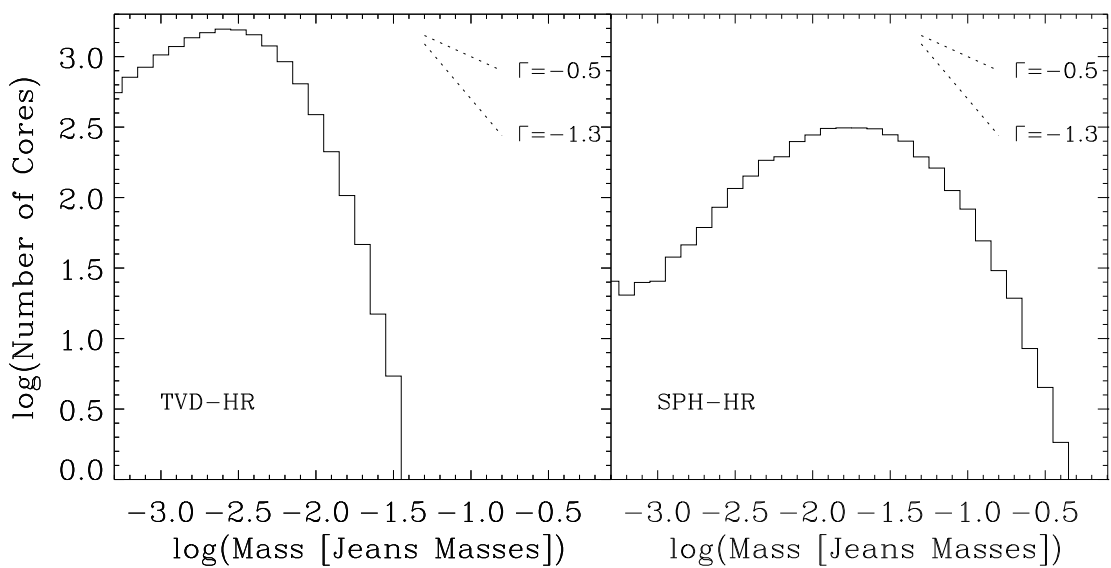}
 \caption{CMD for the high resolution runs using (a) the TVD, and (b)
 SPH schemes. Note that even at high resolution, a single power-law
 not necessarily reproduces the high-mass wing of the CMD.  Instead, a
 function that changes more continuously its slope (probably a
 log-normal function), from zero at maximum, to large negative values
 for increasing masses, may reproduce better the distribution.
\label{cmd_hr:fig}}
\end{figure}

A point of concern is whether our results depend on the resolution of
the simulations.  In Fig.~\ref{cmd_hr:fig} we show the CMD resulting
from the two high-resolution (HR) simulations.  Recall that the HR-TVD
runs has $512^3$ pixels, while the HR-SPH runs has 9,938,375
particles.  This means that the (spatial) resolution is 8 times larger
in the TVD-HR case, while the mass resolution in the SPH-HR case is
almost 50 times larger.  From this figure we stress that the shape of
the CMD is, indeed, not a power-law function, but something more
similar to a log-normal function, for which the slope varies from zero
at the maximum of the distribution to large negative values.

Recall at this point that, since the clumpfinding algorithm in the SPH
runs has been used with a smaller density threshold, the typical core
mass in the SPH models exceeds that in the TVD case.  However, the
general trend of increasing number of cores and decreasing core mass
with increasing RMS Mach number is independent of the used numerical
method as well as the details of the clumpfinding scheme.

\section{Discussion}\label{discussion:sec}

In this section we examine the implications of the results presented in 
\S\ref{results:sec}, in the context of previous work on the CMD 
and its relationship with the stellar IMF.  We find that the CMD is
not a power-law and has a shape that depends on the RMS Mach number.
This fact contradicts previous results stating that, in the magnetic
case, the CMD follows a power-law whose slope depends on the power
index of the kinetic energy spectrum (PN02).  As has been mentioned,
our simulations do not include magnetic fields and this is an
important difference with the work of PN02.  However, this fact is not
the reason for the differences between the shape of CMD reported in
previous section and the power-law they obtain.  Indeed, it can be
shown that a similar analysis to the one performed by PN02 but for the
non magnetic case, also leads to a power-law in which the RMS Mach
number of the flow does not enter in the functional form of $N(m)\,
d\log{m}$.  It only determines the normalization factor, which, in
turn, implies that larger $M_s$ leads to larger number of cores.
Although this fact is consistent with our numerical results (see
\S\ref{results:sec}), our results show that the actual shape of the
CMD does vary with $\Ms$.

We identify the main reason why the model by PN02 does not include the
dependency of the shape of the CMD on $\Ms$ as hinted before: density
fluctuations (cores) in a purely turbulent fluid are built-up by a
statistical superposition of shocks.  This means that mass and density
contrast in individual cores grow by a {\it sequence} of density jumps
with pre-shock densities that can significantly deviate from the mean
value $\rho_0$ (see, e.g., Smith et al.\ 2000).  In addition, in a
turbulent flow with characteristic Mach number $\Ms$, the Mach numbers
of the individual shocks may deviate considerably from the mean value
$\Ms$.  This translates into a large scatter of the Larson (1981)
$\delta v-R$ relationship, which has been observed both in theoretical
works (e.g., V\'azquez-Semadeni et al. 1997; Ostriker et al. 2001;
Ballesteros-Paredes \& Mac Low 2002), as well as in works that include
more recent sets of observations includding different tracers (see
Garay \& Lizano 1999).  Instead, the PN02 model and its hydrodynamic
pendant introduced above, assumes that all cores form in one single
event with fixed Mach number and initial density.  Other caveat in the
PN02 model is that stream collisions form sheets, not cores.  Thus,
the size $\lambda$ of the shock-bounded region may not be a good
representative of the size of cores.

%\subsection{Implications of the turbulent fragmentation process on
%cloud and core substructure}

If cores are generated not by single shocks, but by statistical
superposition of subsequent shocks of the turbulent flow, the
dependency of the density structure on the RMS Mach number may have
important implications for the fragmentation behavior of molecular
clouds, clumps and their cores.  For instance, the internal density
structure of a dense core (even if it is subsonic) may differ
depending whether it is embedded in a highly supersonic, turbulent
molecular cloud, or in a more quiescent one (see, e.g.,
Ballesteros-Paredes et al.\ 2003, Klessen et al.\ 2005).  For high
Mach numbers, the high- frequency part of the turbulent energy cascade
lies in the supersonic regime and, hence, is able to produce strong
density fluctuations on intermediate to small scales.  The amount of
energy at the largest scales, where most of the power is, thus, should
define the fragmentation on all scales, included those where, the
non-thermal motions became subsonic.  Thus, we speculate that cores of
a given mass in more turbulent clouds (e.g., Orion) will have more
substructure than cores of the same mass in less turbulent clouds
(e.g., Taurus).

%AG
%\subsection{Core Mass Spectrum and the IMF}

As mentioned in the introduction, there have been various attempts to
make a direct connection between the CMD and the stellar IMF. However,
the result found in the present work that the CMD has an strong
dependence on the Mach number implies that the relationship between
these two distributions is not immediate. It is thus important to
discus about the possible reasons of this fact in the context of
turbulent fragmentation.

The supersonic turbulent motions ubiquitously observed in interstellar
clouds, suggest that cores form in first place via shocks, with
gravity taking over in the densest and most massive regions.  Once gas
cores become gravitationally unstable, collapse sets in, but the
details of that collapse are not well understood, yet.  It is tempting
to suggest that turbulence plays a critical role in determining the
stellar IMF (see, e.g., the discussion in Ballesteros-Paredes 2005).
In this case, the IMF will directly follow from the CMD if individual
molecular cloud cores map one-to-one onto stars.  However, there is a
large number of physical processes that may play an important role
during protostellar collapse.  Thus, the main problem relating the CMD
with the IMF is that a core with a given mass will not necessarily
form a star of the same mass, or of a mass that is proportional to the
core mass.  However, even considering that these effects do not play
an important role, and that every core of mass $M$ does indeed produce
a star whose mass is directly proportional to $M$, the numerical
results presented in this paper suggest that turbulent fragmentation
is not the sole agent defining the shape of the IMF, since the shape
of the CMD seems not to be a power-law and it varies (as well as the
CMD normalization) with the total amount of kinetic energy.  Thus,
although turbulent fragmentation may play an important role in the
determination of the CMD,
%the lack of a real power-law behavior over one or more orders in mass range makes
it is difficult to argue that it is the main ingredient for
determining the IMF.  In a supersonic turbulent fluid, where the gas
is compressed into filaments and sheets, the initial CMD may depend on
the parameters of the turbulent gas, however there is still room for
other processes to take over and modify the initial CMD in its way to
an intermediate mass distribution, and finally to the IMF.

Finally, it has to be mentioned that, although it is generally
accepted that the IMF has a slope of $-1.3$ at the high-mass range,
the universal character of the IMF is still a debated issue.  For
instance, Scalo (1998) notes that observations taken at face value
reveal variations in the IMF between different star clusters.  Kroupa
(2001; 2002), on the other hand, argues that these variations can be
explained by stellar dynamical processes during the early phases of
star-cluster evolution, like mass segregation and the preferred
evaporation of low-mass stars.  On similar grounds, also Elmegreen
(2000) argues in favor of a more universal IMF, although he recognizes
statistical deviations.

\section{Conclusions}\label{conclusions:sec}

Using two different numerical schemes (TVD and SPH), we have shown
that the core mass distribution (CMD) or core mass spectrum from
purely turbulent (i.e. without self-gravity and magnetic fields)
fragmentation follows a distribution with a continuously varying
slope, not necessarily a single power-law.  The parameters of this
distribution depend on the kinetic energy density in the system as
characterized by the RMS Mach number of the flow.  Interactions of
shocks in high $\Ms$ flows are able to produce strong density
fluctuations on small and intermediate scales.  We thus speculate that
quiescent cores embedded in highly supersonic molecular clouds should
exhibit more internal structure than quiescent cores embedded in more
quiescent clouds.
%The overall density structure is strongly fragmented on
%intermediate to small scales while the occurrence of large-scale
%coherent structures is suppressed.  
As consequence, for a molecular cloud of a given mas, the resulting
resulting CMD is characterized by a large number of low-mass cores
when the Mach number is high.  Low-$\Ms$ flows, on the other hand, are
characterized by small density contrasts and the CMD features fewer
cores with on average larger masses.  

We conclude that the RMS Mach number of the turbulent flow is a very
important parameter that determines the shapes of CMDs.  Our
calculations indicate that the exact shape of the CMD is not
universal, but changes with the Mach number.  This calls the existence
of a simple one-to-one mapping between the purely turbulent CMD and the
observed stellar initial mass function (IMF) into question.  Finding
the true relation between CMD and IMF instead is a very complex and
difficult task, and requires to take additional physical processes
into account, such as protostellar feedback from new-born stars
(bipolar outflows, winds, and radiation), subfragmentation of
individual collapsing protostellar cores into binary and low-$N$
multiple systems, competitive accretion to name just a few.

\acknowledgements

We thank E. V\'azquez-Semadeni and S. Kitsionas for fruitful
discussions.  This work was supported in part by CONACYT grant
27752-E. The work by AG was supported by the DGAPA grant
PAPIIT-IN114802.  J.S.K. was supported by the Astrophysical Research
Center for the Structure and Evolution of the Cosmos (ARCSEC) of the
Korea Science and Engineering Foundation through the Science Research
Center (SRC) program.  R.S.K. and A.K.J. were supported by the Emmy
Noether Program of the DFG under grant KL1358/1.  The TVD simulations
were performed on the Linux clusters at KASI and CRyA, with funding
from KASI and ARCSEC, and CONACYT grant 27752-E, respectively.  The
SPH calculations have been performed at the University of California
at Santa Cruz as well as the PC cluster of the Astrophysical Institute
Potsdam.  This work has made extensive use of NASA's Astrophysics
Abstract Data Service and LANL's astro-ph archives.

%mach 3 => L 0.36 pc
%Mach 6 => L 1.44
%Mach 10 => L=4pc


\begin{thebibliography}{}


\bibitem[Ballesteros-Paredes(2005)]{2005Ap&SS}
Ballesteros-Paredes, J.\ 2005, \apss, in press

\bibitem[Ballesteros-Paredes et al.(2003)]{2003ApJ...592..188B} 
Ballesteros-Paredes, J., Klessen, R.~S., \& V{\' a}zquez-Semadeni, E.\ 
2003, \apj, 592, 188 
 
\bibitem[Ballesteros-Paredes \& Mac Low(2002)]{2002ApJ...570..734B}
Ballesteros-Paredes, J.~\& Mac Low, M.\ 2002, \apj, 570, 734

\bibitem[Ballesteros-Paredes, V{\' a}zquez-Semadeni, \&
Scalo(1999)]{1999ApJ...515..286B} Ballesteros-Paredes, J., V{\'
a}zquez-Semadeni, E., \& Scalo, J.\ 1999, \apj, 515, 286

%\bibitem[Basu \& Jones(2004)]{2004MNRAS.347L..47B} Basu, S., \& Jones,
%C.~E.\ 2004, \mnras, 347, L47

\bibitem[Bate \& Bonnell(2004)]{2004MNRAS.tmp..728B} Bate, M.~R., \&
Bonnell, I.~A.\ 2004, \mnras, 728

\bibitem[Benz 1990]{benz90}
Benz, W. 1990, in The Numerical Modeling of Nonlinear Stellar Pulsations,
ed. J. R. Buchler (Dordrecht: Kluwer), 269

%\bibitem[Blitz(1993)]{1993prpl.conf..125B} Blitz, L.\ 1993, Protostars and
%Planets III, 125

\bibitem[Chabrier(2003)]{2003PASP..115..763C} Chabrier, G.\ 2003, \pasp,
115, 763

\bibitem[Elmegreen (1993)]{elme93}Elmegreen, B. G. 1993, \apj, 419, L29

\bibitem[Elmegreen(2000)]{2000ApJ...539..342E} Elmegreen, B.~G.\ 2000, 
\apj, 539, 342 

\bibitem[Gammie, Lin, Stone, \& Ostriker(2003)]{2003ApJ...592..203G}
Gammie, C.~F., Lin, Y., Stone, J.~M., \& Ostriker, E.~C.\ 2003, \apj, 592,
203

\bibitem[Garay \& Lizano(1999)]{1999PASP..111.1049G} Garay, G., \& Lizano, 
S.\ 1999, \pasp, 111, 1049 
 
\bibitem[Goodwin et al.(2004)]{2004A&A...423..169G} Goodwin, S.~P.,
Whitworth, A.~P., \& Ward-Thompson, D.\ 2004, \aap, 423, 169

\bibitem[Hartmann(2001)]{2001AJ....121.1030H} Hartmann, L.\ 2001, \aj, 121, 
1030 
 
\bibitem[Hunter \& Fleck (1982)]{hunter82} Hunter, J. H., \& Fleck, R. C.
1982, \apj, 256, 505

\bibitem[Jappsen et al. 2005]{jappsen_etal05} Jappsen A. K., Klessen R.~S.,
Larson, R.~B., Li Y., \&  Mac Low, M.~M., 2005, \aap, 435, 611

\bibitem[Kim et al. 1999]{1999ApJ...514..506K}
Kim, J., Ryu, D., Jones, T.~W., \& Hong, S.~S. 1999, \apj, 514, 506

\bibitem[Kim et al. 2005]{Kim_etal05}Kim, J.S. et al. 2005. \apj, submitted

%\bibitem[Klessen (2000)]{2000ApJ...535..869} Klessen, R.~S. 2000
%\apj, 535, 869

\bibitem[Klessen (2001)]{2001ApJ...556..837K}
 Klessen, R.~S. 2001, \apj, 556, 837

\bibitem[Klessen et al. (2005)]{Klessen_etal05} Klessen, R.~S.,
Ballesteros-Paredes, J., V{\' a}zquez-Semadeni, E., \& Dur\'an-Rojas,
C. 2004, \apj, 620,786

\bibitem[Klessen \& Burkert(2000)]{2000ApJS..128..287K} Klessen, R.~S., \& 
Burkert, A.\ 2000, \apjs, 128, 287
 
\bibitem[Klessen et al.(2000)]{2000ApJ...535..887K} Klessen, R.~S.,
Heitsch, F., \& Mac Low, M.\ 2000, \apj, 535, 887

\bibitem[Klessen \& Lin (2003)]{Klessen_lin03} Klessen, R.~S., \& Lin, D.N.C.
PRE, 67, 036311

\bibitem[Kroupa(2001)]{2001MNRAS.322..231K} Kroupa, P.\ 2001, \mnras, 322, 
231

\bibitem[Kroupa(2002)]{2002ASPC..285...86K} Kroupa, P.\ 2002, Astronomical
Society of the Pacific Conference Series, 285, 86

\bibitem[Larson(1981)]{1981MNRAS.194..809L} Larson, R.~B.\ 1981, \mnras,
194, 809

\bibitem[Larson 1985]{1985MNRAS.214..379L}
Larson, R. B. 1985, MNRAS, 214, 379

\bibitem[Larson(2005)]{2005MNRAS.359..211L} Larson, R.~B.\ 2005, \mnras, 
359, 211 

\bibitem[Li, Klessen \& Mac Low (2003)]{2003ApJ...592..975L}Li, Y.,
Klessen, R. S., \& Mac Low, M.~M. 2003, \apj, 592, 233L

\bibitem[Li et al.(2004)]{2004ApJ...605..800L} Li, P.~S., Norman, M.~L., 
Mac Low, M., \& Heitsch, F.\ 2004, \apj, 605, 800 
 
%\bibitem[Li \& Nakamura(2004)]{2004ApJ...609L..83L} Li, Z.~\& Nakamura, F.\
%2004, \apjl, 609, L83

%\bibitem[Mac Low, Klessen, Burkert, \& Smith(1998)]{1998PhRvL..80.2754M}
%Mac Low, M., Klessen, R.~S., Burkert, A., \& Smith, M.~D.\ 1998, Physical
%Review Letters, 80, 2754

\bibitem[Mac Low \& Klessen (2004)]{2004RvMP...76..125M}
Mac Low, M., Klessen, R.~S. 2004, RvMP, 76, 125

\bibitem[Meyer et al. (2000)]{Meyer_etal00} Meyer, M. R., Adams, F.C.,
Hillenbrand, L.A., Carpenter, J.M., \& Larson, R.B. 2000, in
Protostars and Planets IV, ed. V. Mannings, A. P. Boss, \&
S. S. Russell (Tuscon: Univ. Arizona Press), 121


\bibitem[Motte et al. (1998)]{1998A&A...336..150M}Motte, F., Andr\'e, P.,
\& Neri, R. 1998, \aap, 336, 150

\bibitem[Ossenkopf et al.(2001)]{2001A&A...379.1005O} Ossenkopf, V., 
Klessen, R.~S., \& Heitsch, F.\ 2001, \aap, 379, 1005 
 
\bibitem[Ostriker, Stone, \& Gammie(2001)]{2001ApJ...546..980O} Ostriker,
E.~C., Stone, J.~M., \& Gammie, C.~F.\ 2001, \apj, 546, 980

\bibitem[Padoan, Juvela, Goodman, \& Nordlund(2001)]{2001ApJ...553..227P}
Padoan, P., Juvela, M., Goodman, A.~A., \& Nordlund, {\AA}.\ 2001b, \apj,
553, 227

\bibitem[Padoan \& Nordlund(2002)]{2002ApJ...576..870P} Padoan, P., \& 
Nordlund, {\AA}.\ 2002, \apj, 576, 870 
 
\bibitem[Passot \& Vazquez-Semadeni (1998)]{1998PhRvE..58.4501P}
Passot, T., \& V\'azquez-Semadeni, E. 1998, PhRvE, 58, 4501

\bibitem[Salpeter (1955)]{1955ApJ...121..161S}
Salpeter, E.~E. 1955, \apj, 121, 161

\bibitem[Sasao 1973]{sasao}Sasao, T. 1973, PASJ, 25, 1

\bibitem[Scalo (1998)]{1998ASPC..142..201S}
Scalo, J. 1998, in The Stellar Initial Mass Function (38th
Herstmonceux Conference), G. Gilmore \& D. Howell eds. ASPC, 142, 201

\bibitem[Scalo \& Elmegreen (2004)]{2004ARA&A..42..275S}
Scalo, J., \& Elmegreen, B. G. 2004, ARA\&A, 24,275

\bibitem[Scalo et al. 1998]{1998ApJ...504..835S} Scalo, J.,
V\'azquez-Semadeni, E., Chappell, D., \& Passot T. 1998, \apj, 504,
835

\bibitem[Schmeja \& Klessen (2004)]{2004A&A...419..405S}Schmeja, S.,
\& Klessen, R.~S. 2004, \aap, 419, 405

\bibitem[Smith et al.(2000)]{2000A&A...362..333S} Smith, M.~D., Mac Low, 
M.-M., \& Heitsch, F.\ 2000, \aap, 362, 333 
 
\bibitem[Spitzer (1978)]{1978ppim.book.....S} Spitzer,
L. 1978. Physical processes in the interstellar medium, New York
Wiley-Interscience 

\bibitem[Springel et al.(2001)]{2001NewA....6...79S} Springel, V., Yoshida, 
N., \& White, S.~D.~M.\ 2001, New Astronomy, 6, 79 

\bibitem[Stone, Ostriker, \& Gammie(1998)]{1998ApJ...508L..99S} Stone,
J.~M., Ostriker, E.~C., \& Gammie, C.~F.\ 1998, \apjl, 508, L99

\bibitem[Tilley \& Pudritz (2004)]{2004MNRAS.353..769T}
Tilley, D. A., \& Pudritz, R. E. 2004, \mnras, 353, 769

\bibitem[Testi \& Sargent 1998]{1998ApJ...508L..91T}Testi, L., \&
Sargent, A. I. 1998, \apj, 508, 91

\bibitem[Vazquez-Semadeni (1994)]{1994ApJ...423..681V}
V\'azquez-Semadeni, E. 1994, \apj, 423, 681

\bibitem[Vazquez-Semadeni et al.(1997)]{1997ApJ...474..292V} 
V\'azquez-Semadeni, E., Ballesteros-Paredes, J., \& Rodriguez, L.~F.\
1997, \apj, 474, 292 
 
\bibitem[Vazquez-Semadeni et al.(2000)]{2000prpl.conf....3V}
V\'azquez-Semadeni, E., Ostriker, E.~C., Passot, T., Gammie, C.~F., \&
Stone, J.~M.\ 2000, Protostars and Planets IV, 3

\bibitem[Williams et al.(1994)]{1994ApJ...428..693W} Williams, J.~P., de
Geus, E.~J., \& Blitz, L.\ 1994, \apj, 428, 693

\end{thebibliography}
\end{document}